\documentclass[12pt,draftclsnofoot,twocolumn,oneside]{IEEEtran}

\usepackage{listings}
\usepackage{amsmath}
\usepackage{graphicx}
\usepackage{amsfonts}
\usepackage{array}
\usepackage{amssymb}
\usepackage{helvet}
\usepackage{color}
\usepackage{colortbl}
\usepackage{rawfonts,amsfonts,psfrag,flushend,geometry}
\usepackage{bm}
\usepackage{float}
\usepackage{cite}
\usepackage{multirow}
\usepackage{slashbox}
\usepackage{setspace}
\usepackage{times}

\makeatletter
\def\hlinewd#1{%
  \noalign{\ifnum0=`}\fi\hrule \@height #1 \futurelet
   \reserved@a\@xhline}
\makeatother
\definecolor{light-gray}{gray}{0.90}

\begin{document}
\begin{spacing}{0.85}
\title{\huge{Hardware-Constrained Millimeter Wave Systems for 5G: Challenges, Opportunities, and Solutions} }
\author{Xi Yang, Michail Matthaiou, Jie Yang, Chao-Kai Wen, Feifei Gao, and Shi Jin
}

\maketitle

\begin{abstract}
Although millimeter wave (mmWave) systems promise to offer larger bandwidth and unprecedented peak data rates, their practical implementation faces several hardware challenges compared to sub-6\,GHz communication systems. These hardware constraints can seriously undermine the performance and deployment progress of mmWave systems and, thus, necessitate disruptive solutions in the cross-design of analog and digital modules.
In this article, we discuss the importance of different hardware constraints and propose a novel system architecture, which is able to release these hardware constraints while achieving better performance for future millimeter wave communication systems. The characteristics of the proposed architecture are articulated in detail, and a representative example is provided to demonstrate its validity and efficacy.
\end{abstract}

\IEEEpeerreviewmaketitle

\section{Introduction}
Millimeter-wave (mmWave) systems avail of extremely large bandwidth and can, therefore, boost the peak data rates and improve the user experience. For this reason, they have been identified as a catalyst for the next generation wireless communication systems \cite{Intro_mmWave}.

In order to speed up the successful roll-out of mmWave systems and promote the evolution of mobile Internet, a great amount of research has been reported in the literature including channel modeling \cite{Channel_Model}, design of air interface and optimal transceiver algorithms \cite{Air_Interface,Channel_Estimation}, beam training and tracking \cite{Channel_Estimation}, system prototyping \cite{Prototype_mmWave}, etc. Compared with sub-6\,GHz wireless systems, communication over mmWave frequencies experiences substantially different characteristics, such as higher path-loss and diffraction losses, stronger directionality, and larger bandwidth. Yet, the main challenge pertaining to the implementation of mmWave systems is the exacerbated impact of hardware constraints compared to current radio frequency (RF)-based systems. For example, a higher carrier frequency induces larger multiplication factors, which in turn, amplify the phase noise and necessitate the development of robust mitigation algorithms. The increased path-loss needs to be compensated by deploying a high (or even a massive) number of antennas, which will be forming a lens topology or be connected to a bank of phase shifters, in order to maintain the hardware cost and power consumption at reasonable levels. In addition, a stronger directionality means mm-wave systems are sensitive to angular mismatches and demand high-quality analog components. Finally, the higher penetration losses at mmWave frequencies imply that beams are more prone to blockage; a phenomenon that can be mitigated by efficient beam scanning schemes.

\begin{figure*}[b]
  \centering
    \includegraphics[width=140mm]{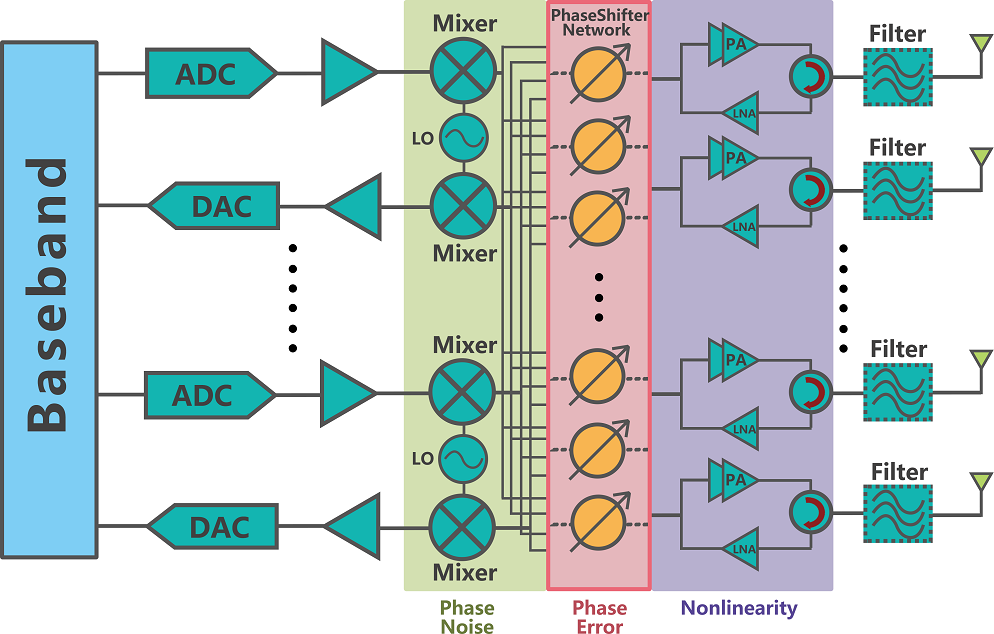}
  \caption{A simplified block diagram of a large-scale antenna array-based mmWave transceiver.}
  \label{P1}
\end{figure*}

As an example, Fig.\,\ref{P1} shows a simplified general block diagram of a large-scale antenna array-based mmWave transceiver. As demonstrated in Fig.\,\ref{P1}, the hardware architecture of a large-scale antenna array-based mmWave transceiver is complicated and contains many RF analog components. More importantly, these analog components are generally imperfect due to manufacturing errors and introduce different hardware imperfections such as phase noise, power amplifier nonlinearities, IQ imbalance and so on, especially in mmWave frequencies. On top of these issues, with the usage of large bandwidth, mmWave systems also encounter challenges of the huge amount real-time baseband signal processing. For these reasons, the performance of mmWave systems may be severely affected by these hardware imperfections and constraints and a thorough analysis and design is of utmost importance along with potential solutions.

In this article, we propose a novel system architecture which leverages dynamic hardware and a bidirectional channel that connects hardware and software, for the purpose of relaxing the above mentioned hardware constraints while achieving better performance for future mmWave communication systems. The remainder of the article is organized as follows. {{We first present an overview of hardware constraints in mmWave systems in Section II. Section III provides challenges and solutions including the proposed novel system architecture. We then outline the future open research topics for hardware-constrained mmWave systems in Section IV. After that, we conclude this article in Section V.}}

\section{Hardware Constraints in mmWave Systems}
In the following, according to the general block diagram of a large-scale antenna array-based mmWave transceiver presented in Fig.\,\ref{P1}, we provide an overview of common hardware constraints that very often appear in mmWave systems.

\subsection{Phase Noise}
Oscillators are important components in wireless communication systems, which along with frequency synthesizers, are responsible for generating clock signals for timing and frequency synchronization, and for providing carrier signals to the mixers in the transmitter and receiver devices. Unfortunately, both oscillators and frequency synthesizers used in communication systems are generally imperfect, since the phase of the clock signals or carrier signals is affected by a broad array of factors, such as phase jitter from the reference oscillator, thermal noise from the phase-locked loop (PLL) circuits, and interference signals coupling from the power supply.

As a result of these factors, a typical output frequency spectrum of a non-ideal oscillator usually has two types of distortions i.e. spurious tones and phase noise. Spurious tones appear as distinct components in the spectrum and belong to long-term instabilities. In particular, spurious tones are slow variations that occur over hours or more, and, therefore, have less impact on the performance. On the contrary, phase noise is identified as a short-term instability, and causes time jitter and reciprocal mixing. In this article, we neglect the imperfections from spurious tones and mainly focus on phase noise. An example of phase noise characteristics is provided in Fig. \ref{P2}(a), which shows how the spectrum spread occurs at the output frequency spectrum of a non-ideal oscillator due to the phase noise. The vertical axis represents the phase-noise spectral density (PSD), which is defined as the ratio of the noise in a 1-Hz bandwidth at a specified frequency offset over the oscillator's signal amplitude at the carrier frequency. Note that the phase noise can be regarded as the characterization of the output of a non-ideal oscillator in the frequency domain, whilst time jitter is the counterpart in the time domain.

The impact caused by phase noise in practical communication system can be summarized as follows: (i) Phase noise is firstly added into the local oscillator (LO), and subsequently conveyed to the generated local carrier signal. Then, the phase noise will be introduced into the received baseband signal through down-conversion by mixing the received RF signal with the impaired local carrier signal. Consequently, the signal-to-noise ratio (SNR) is degraded and the utilization of higher order modulation schemes is substantially compromised. (ii) The time jitter related to phase noise causes aperture uncertainty at the analog-to-digital converters (ADC), which increases the system noise and contributes to the uncertainty in the actual phase of the sampled signal. (iii) Most importantly, the random phase jitter in the time domain resulting from (i) and (ii) will limit the channel coherence time, especially at mmWave frequencies. Since PSD at higher frequency means faster random phase changes and phase noise increases 6\,dB with every frequency doubling, mmWave systems are much more prone to phase noise. Hence, compared with sub-6\,GHz wireless systems, in practical mmWave systems, the channel estimate obtained through pilot symbols is more likely to be corrupted and cannot be used for data recovery. Therefore, it is of great importance to propose phase noise-robust algorithms or design specified frame structure schemes for phase noise impaired mmWave systems.

As a side remark, phase noise is usually modeled as a discrete-time Wiener process or a Gaussian process with specified power spectral density e.g. IEEE Model \cite{Phase_Noise}. Additionally, when combined with large-scale antenna arrays in mmWave systems, different oscillator configurations, i.e., common oscillator (CO) setup, where all antennas are connected to a single oscillator, or distributed oscillator (DO) setup, where each antenna has its own independent oscillator, should also be considered comparatively \cite{E_CoDo}.

\subsection{Nonlinear Power Amplifiers}
Power amplifiers (PAs) are responsible for boosting the transmit power to overcome the path loss between the transmitter and the receiver. Ideally, an ideal amplifier should exhibit linear transfer characteristics and output a replica of the input multiplied by a scalar. However, to achieve a better power efficiency, practical amplifiers always suffer from various nonlinearities, and these nonlinearities can be categorized into two categories. The first one is referred as amplitude modulation/amplitude modulation (AM/AM) conversion, which describes the relationship between the output power and input power, and the latter one is referred as amplitude modulation/phase modulation (AM/PM) conversion, which illustrates the relationship between the output phase and input power.

\begin{figure*}[b]
  \centering
    \includegraphics[width=150mm]{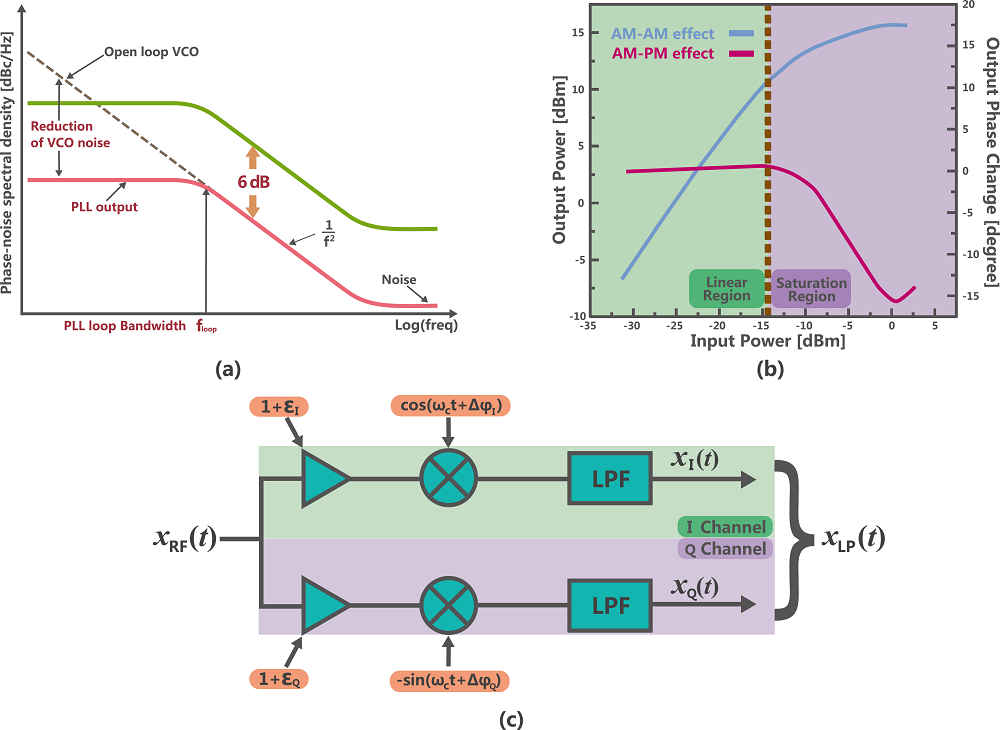}
  \caption{Visualization of different hardware imperfections {{in a mmWave system}}: (a) An example of phase noise characteristics. (b) The operation characteristics of a practical power amplifier. (c) A sketch of the principle of IQ-imbalance.}
  \label{P2}
\end{figure*}

The operational characteristics of a practical power amplifier are shown in Fig. \ref{P2}(b). As demonstrated in Fig. \ref{P2}(b), the practical power amplifier has different operating regions i.e. linear region and saturation region. In the linear region, the power amplifier exhibits linear transfer characteristics, whilst the power efficiency of the amplifier is not optimal. When the input power increases, the power amplifier goes into the saturation region, where its power efficiency is maximized but, unfortunately, nonlinearities start to appear. These nonlinearities cause amplitude distortions and phase distortions of the transmitted signal and induce interference to adjacent frequency tones due to the spectral spreading of the distorted signals.

In order to mitigate these nonlinearities, output backoff (OBO) is adopted in most current power amplifiers, which reduces the transmit power and keeps the power amplifier operating within its linear region. However, in this way, the maximum transmit power is inherently reduced and the amplifier efficiency is substantially decreased, especially in mmWave systems. Another way to alleviate the nonlinearities is through digital pre-distortion (DPD), which means that we pre-process the transmitted signal in the digital domain based on the captured nonlinear transfer characteristics of the practical power amplifier, so that we can obtain approximately linear characteristics.
Nonetheless, the drawback of DPD is its rapidly increasing computation complexity with the number of antennas.
Additionally, with the adoption of ultra-large bandwidth, the computation complexity of the DPD will be further increased and may even suspend its utilization.
Note that there are two categories of power amplifier models{{\cite{Power_Amplifier}}} i.e. frequency-independent models and frequency-dependent models. Frequency-independent models{{\cite{Power_Amplifier}}} include the polynomial model, Rapp model and white model, which fail to effectively capture frequency dependent distortion. On the other hand, frequency-dependent models{{\cite{Power_Amplifier}}} include the Saleh model and Wiener model, and are suitable for modeling nonlinearities in wideband systems.

\subsection{IQ-Imbalance}
Quadrature sampling is dominant in modern communication transceivers, and is anticipated to be utilized in mmWave systems as well. Nevertheless, due to manufacturing imperfections, the in-phase (I) branch and quadrature-phase (Q) branch employed for quadrature sampling are unlikely to be perfectly matched. The phase difference between the output I and Q signals from the local oscillator splitter is never exactly $90^o$, and the amplitude gain of I and Q signals are also not consistent as well because of the amplitude errors in digital-to-analog converters (DACs) or inconsistencies in the analog mixers. Thus, IQ-imbalance kicks in and an attenuated and rotated version of the original signal appears as shown in Fig. \ref{P2}(c), where $\varepsilon_I$ and $\Delta\varphi_I$ represent the amplitude and phase offsets in the I branch, and $\varepsilon_Q$ and $\Delta\varphi_Q$ represent the amplitude and phase offsets in the Q branch, respectively.

For mmWave systems, IQ-imbalance is more exacerbated in comparison with sub-6\,GHz wireless systems due to higher operating frequency and larger bandwidth. Besides that, IQ-imbalance may also impact different large-scale antenna array architectures in mmWave, i.e., full digital beamforming and hybrid beamforming architectures \cite{Channel_Estimation}.

Two types of IQ-imbalance models are usually considered in IQ-imbalance compensation algorithmic design and performance analysis, i.e., frequency-independent and frequency-selective models \cite{IQ_Imbalance}. Frequency-independent models only consider quasi-linear impairments of the input signal, while frequency-selective model capture the behaviour of the analog components more accurately.

\subsection{Phase Shifters vs Lens Topologies}
The adoption of beamforming in mmWave systems, for the purpose of compensating the severe path loss and improving the effective signal-to-interference ratio (SINR), necessitates the usage of large-scale phased antenna arrays or lens topologies \cite{Channel_Estimation}.

As crucial components in any large-scale phased antenna array, phase shifters have attracted considerable interest recently. Generally, there are two basic types of phase shifters for mmWave systems i.e., active phase shifters and passive phase shifters. By making efficient use of active modules, the former outperform their passive counterparts in terms of insertion loss and maximize the spectral efficiency. On the other hand, the passive phase shifters consume nearly zero power and can deliver better energy efficiency. Both types of phase shifters are costly (may contribute to one-third of the cost of phased antenna arrays) and experience phase errors due to manufacturing tolerances and material imperfections  \cite{Phase_shifter}. These phase errors include deterministic errors and random errors. The former can be corrected through appropriate manufacturing while the latter can have a multiplicative effect on the beam pattern and need to be compensated algorithmically. For example, in a 6-bit digital phase shifter system, the phase accuracy of phase shifters is limited and the actual shifted phase could be uniformly distributed between the minimum phase interval, i.e. $5.6^o$, at each specified stage.Such random phase errors may destroy the beam pattern and induce interference especially in multi-user communications.

In comparison with phased antenna arrays, lens-based antenna arrays can provide wider bandwidth and directional narrow beams with much lower cost and complexity, especially in multi-beam scenarios. Unlike phased antenna arrays, lens topologies can provide multiple beams simultaneously by simply deploying multiple feeders instead of configuring a complicated beam-forming feed network and phase shifter drivers. Nevertheless, lens-based antenna arrays still suffer from manufacturing imperfections while their physical dimensions are inherently large. The loss of focus in lens-based antenna array, due to the limited number of arranged feeders around the lens, is also a great challenge.

\subsection{Ultra-High Data Processing Pressure}
By leveraging the GHz transmission bandwidth and sampling rate, the data throughput in mmWave systems increases drastically. For example, for a 2\,GHz bandwidth single-input single-output single-carrier mmWave system, with quadrature sampling rate of 3.072\,GSample/s and ADC resolution of 12\,bit, the data throughput is up to 73\,Gbps. Furthermore, when combined with a large-scale antenna array and multiple RF chains, the data throughput would theoretically be hundreds of Gbps. Such unprecedented data rates not only create formidable challenges on the data exchange interface between the remote radio unit and the base station baseband unit, but also put enormous pressure on the design and implementation of baseband algorithms, which are responsible for digital signal processing under very strict time constraints.

\section{Challenges, Opportunities, and Solutions}
All the above mentioned hardware constraints can be roughly summarized into two categories, i.e., conventional hardware constraints and mmWave-encountered hardware constraints. Conventional hardware constraints include phase noise, nonlinearities from power amplifiers, and IQ-imbalance and so on. All these conventional hardware constraints also exist in sub-6\,GHz wireless systems but their impact is exacerbated in mmWave systems. On the other hand, mmWave-encountered hardware constraints contain the ultra-high data processing pressure and the imperfections resulted from the components of the wideband beamforming hardware, e.g., phase shifters or lens arrays, etc. Moreover, the low power and low cost design is also a great challenge in future mmWave system architecture.

\subsection{Compensation of Conventional Hardware Imperfections}
To compensate these conventional hardware constraints, a theoretical analysis should be firstly performed with the objective of determining what their effects are and what is the optimal hardware quality to achieve predetermined performance targets. Then, as a next step, advances can be made in both hardware or software. For example, with respect to phase noise, we can suppress the phase noise levels by the usage of more advanced oscillators with better PLL which are very expensive, or we can only utilize sophisticated algorithms that are robust to phase noise, depending on the beamforming performance we plan to achieve. In addition, the characteristics of mmWave systems e.g., channel sparsity, should be efficiently exploited in the design of compensation algorithms by taking into account the large-scale antenna arrays and wide bandwidth.

\subsection{Solutions for mmWave-encountered Hardware Constraints}
The high cost of phase shifter networks and the performance degradation induced by random phase errors may be addressed by integrating few high-precision phase shifters into the mixed beamforming network. The focus loss of lens topologies can be alleviated by the combined use of lens and phase shifters, where the lens is used for beamforming while the phase shifters are used for multiple adjacent beams combining. Additionally, an analog beamforming network could also be leveraged, such as a Butler matrix \cite{Butler}, to simplify the complicated RF front ends. Regarding the stringent high data-processing requirements, low-resolution ADCs in hardware combined with advanced receiver algorithms, e.g. the generalized expectation consistent (GEC) signal recovery algorithm \cite{GEC}, emerges as a viable candidate.

However, it is still difficult to establish a general framework embracing all these hardware constraints and imperfections. Most importantly, it is immensely difficult to achieve optimal performance for the entire system by optimizing different modules separately. Recall that there are still multiple barriers in designing optimal transceiver algorithms for conventional architectures, where the hardware always has less flexibility and the baseband operation is inherently bonded by the fixed hardware architecture \cite{Channel_Estimation}. Both these observations imply that it is very demanding to address all hardware and software challenges appearing in mmWave systems. Hence, in the next subsection, we propose a hardware-aided system architecture for future mmWave systems which avails of a dynamic hardware structure and relaxes the hardware constraints at the cost of additional hardware complexities.

\subsection{Proposed Hardware-Aided System Architecture for Future mmWave Systems}
\begin{figure*}[b]
  \centering
    \includegraphics[width=1.0\textwidth]{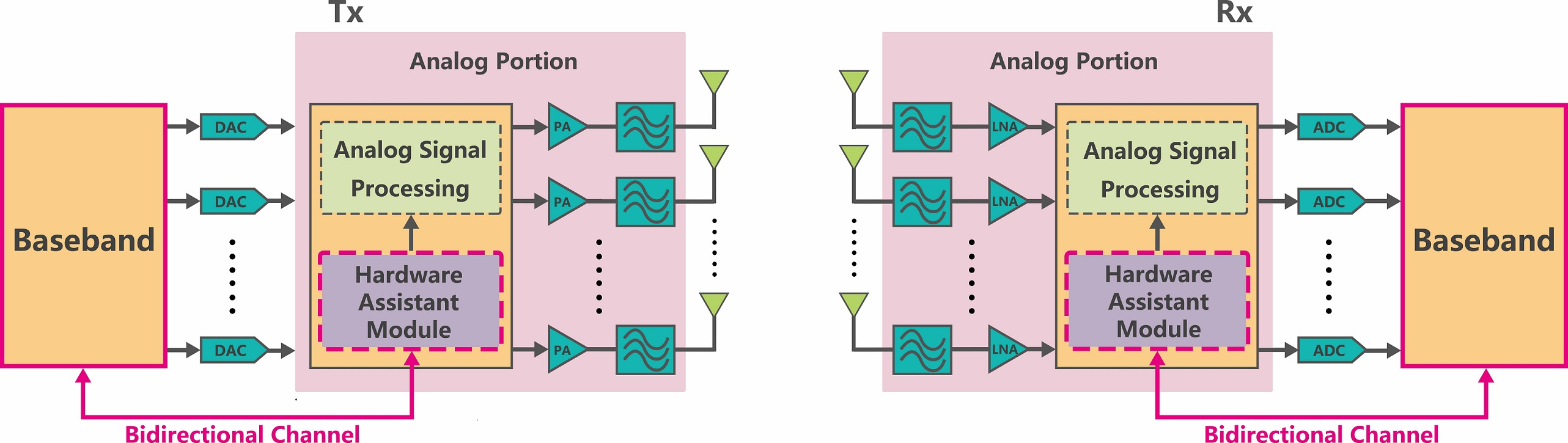}
  \caption{The block diagram of the proposed dynamic hardware-aided system architecture for future mmWave systems.}
  \label{P3}
\end{figure*}

Fig.\,\ref{P3} shows the generalized proposed hardware-aided system architecture for future mmWave systems, where the analog signal processing module and the hardware assistant module are employed in the analog portions. The analog signal processing module is generally made up of various phase shifter networks or Butler circuits, and its responsibility is to perform analog-favored pre-processing, such as signal combining and beamspace transformation, to effectively decrease the data exchange between the analog and digital interfaces.

The design purpose of the hardware assistant module is to assist the analog signal processing module in specified application requirements e.g., energy detection in multi-beam scenarios, analog beamforming coefficients' calculation and calibration for accurate beam pattern control, adaptive phase shifter network structure adjustment under hybrid beamforming architecture, and interference cancellation in full duplex mode and so on. Therefore, the detailed implementation of the hardware assistant module varies depending on the application requirements. For example, in lens-based mmWave systems, the hardware assistant module can be an energy detection circuit and be responsible for multi-beam energy detection, and thus help the analog signal processing module perform efficient signal combining and eliminate the focus loss problem. Furthermore, the hardware assistant module can also be a low-resolution ADC-based full digital RF front end to help achieve fast beam training in mmWave systems, as demonstrated in\cite{Yang18}.

A bidirectional channel is also introduced to exchange real-time information between the baseband unit and the hardware assistant module, such as control commands from the baseband or the number of beams detected from the analog portion.

Compared with conventional system architectures, the proposed hardware-aided system architecture offers the following advantages: 1) The analog unit is enhanced by introducing analog signal processing and hardware assistant modules; thus, ultra-high throughput data can be pre-processed as expected in the analog domain, e.g., through Discrete Fourier Transform (DFT) or signal combining, which will effectively relieve the analog-digital interface from the ever increasing data processing requirements. 2) By cooperating with the hardware assistant module, the RF front ends of mmWave systems can avail of better hardware flexibility and be resilient to hardware imperfections. Moreover, some useful assistant information can also be acquired timely and conveniently.

\begin{figure*}[b]
  \centering
    \includegraphics[width=1.0\textwidth]{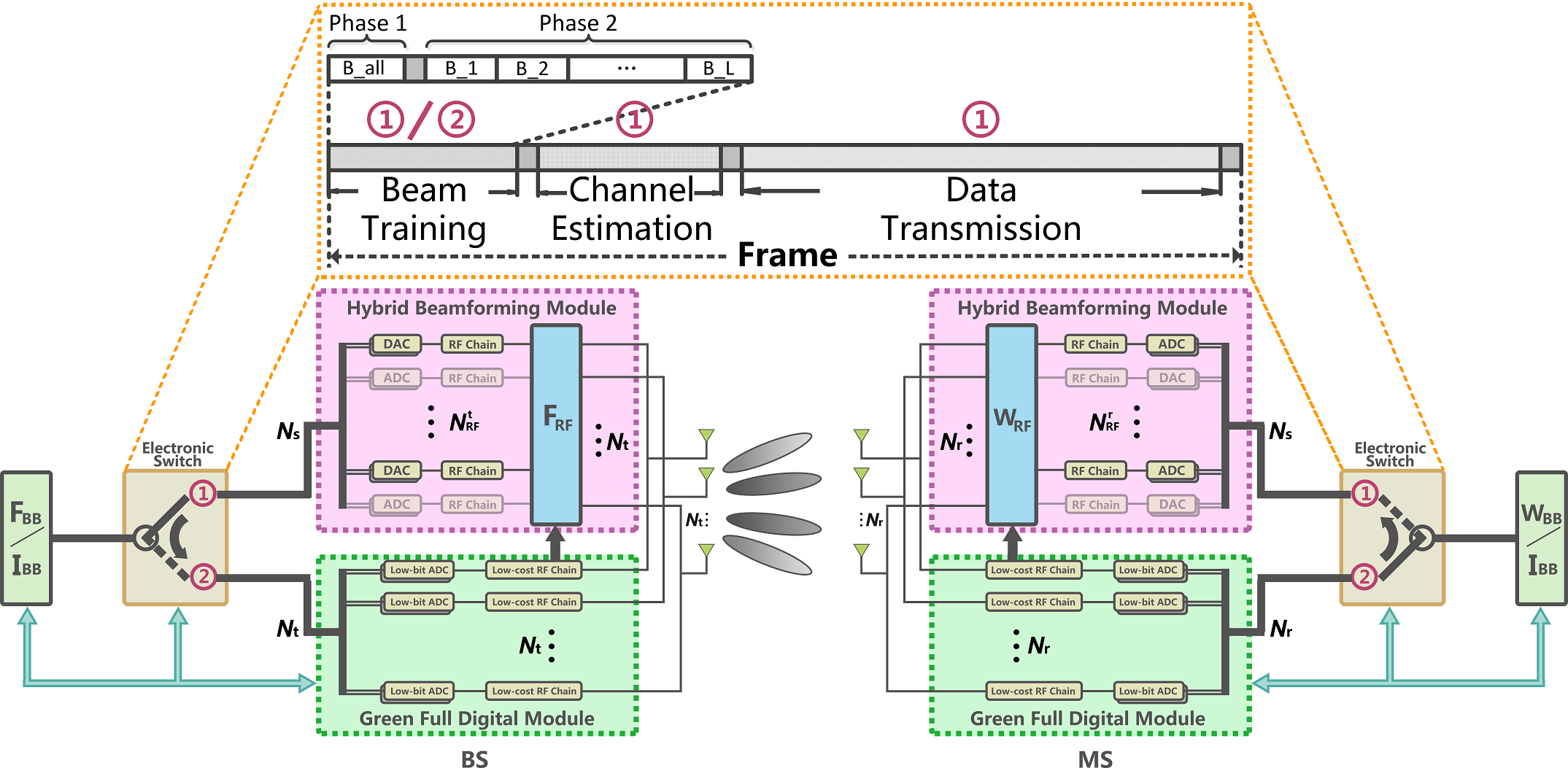}
  \caption{{A representative example of the proposed hardware-aided system architecture for hybrid mmWave systems.}}
  \label{P4}
\end{figure*}

To demonstrate the validity and efficacy of the proposed system architecture, we describe a representative example to showcase how the assistant hardware could be used to alleviate beam training overhead in a hybrid mmWave system in\cite{Yang18}. The novel mmWave transceiver with the capability of fast beam training is presented in Fig.\,\ref{P4}.
In comparison with the conventional hybrid architecture in \cite{Training_Overhead}, whose training overhead is ${K^2}{\log _K}{G_t}$ slots where $K$ is the number of sectors and $G_t$ stands for the number of transmit antennas, the overhead needed for our proposed architecture is only $L+1$ slots, where $L$ represents the number of paths in the mmWave finite dimensional channel model of \cite{Channel_Estimation}. In other words, the training overhead for our proposed architecture only scales with the actual beam directions in the propagation environment.

\section{Future Research Directions}
In the following, we outline the future open research topics for hardware-constrained mmWave systems from aspects of advanced array signal processing, baseband signal processing algorithm, and advanced hardware architecture in terms of the requirements of the ultra-high baseband data processing.

\subsection{mmWave with Advanced Array Signal Processing}
To overcome the large path-loss, mmWave systems are typically configured with large-scale antenna arrays at the base station. Hence, exploiting advanced array signal processing in mmWave systems is important for the purpose of reducing signal processing complexity and, at the same time, achieving enhanced performance. Particularly, the inherent array structure of massive antennas in mmWave systems could be re-conceptualized and explored from the array signal processing viewpoint, e.g., direction of arrival/departure and beam squint, which can be then applied in mmWave system operation design, such as channel estimation, hybrid beamforming, interference control, multiple access scheme, and mobility management via angle tracking.

\subsection{Machine Learning-Aided Baseband Signal Processing Algorithm}
Recently, machine learning (ML), especially deep learning (DL), has been successfully applied for improving the performance of baseband signal processing algorithms\cite{Wang17}, including channel decoding, channel estimation and detection, and so on. Therefore, to achieve better performance in mmWave systems, considering the difficulties and complexities of deducing an accurate analytical signal model that embraces all the hardware constraints and imperfections, it could be more practical and of great benefit to leverage DL. By using the training dataset (pre-obtained or online) to learn the actual signal model and refine the parameters of algorithms via DL, mmWave systems may be able to overcome hardware constraints and imperfections in terms of signal processing algorithms.

\subsection{Advanced Hardware Architecture}
Given the ultra-high data processing pressure in mmWave systems, it is significant to develop advanced hardware architectures from two aspects i.e., data rate reduction and high-speed parallel data processing. On one hand, despite the deployment of low-resolution ADCs in RF front ends, channel sparsity and narrow angular spread should also be exploited in hardware architecture design to reduce baseband data processing dimension for mmWave systems. On the other hand, it seems promising to implement high-speed parallel baseband data processing by means of field programmable gate array (FPGA)-assisted multi-core general purpose processor-based baseband architectures when combined with carrier aggregation.

\section{Conclusion}
In this article, an overview of hardware constraints has been provided and the impact of hardware imperfections on mmWave systems has been analyzed. In order to relax the hardware constraints, we proposed a hardware-aided system architecture, which leverages the consistent interaction between hardware and software units and offers better flexibility and adaptability compared to the state of the art. An indicative example demonstrated the superiority of the proposed solution. Our future research will include the physical design of the hardware assistant module.

\section*{Acknowledgements}
This work was supported in part by the National Science Foundation (NSFC) for Distinguished Young Scholars of China with Grant 61625106, and in part by the National Natural Science Foundation of China under Grants 61531011 and 61831013. The work of M. Matthaiou was supported by EPSRC, UK, under Grant EP/P000673/1. The work of C.-K. Wen was supported by the Ministry of Science and Technology of Taiwan under Grants MOST 107-2221-E-110-026 and the ITRI in Hsinchu, Taiwan.

\section*{Biographies}
Xi Yang (ouyangxi@seu.edu.cn) received the the B.S. degree and the M.S. degree from Southeast University, Nanjing, China in 2013 and 2016 respectively. She is currently working toward the Ph.D. degree with the School of Information Science and Engineering, Southeast University. Her main research interests include massive MIMO system prototyping, beam training and channel estimation in millimeter wave, and millimeter wave system prototyping.

{Michail Matthaiou} [S'05-M'08-SM'13] (m.matthaiou@qub.ac.uk) was born in Thessaloniki, Greece in 1981. He obtained the Diploma degree (5 years) in Electrical and Computer Engineering from the Aristotle University of Thessaloniki, Greece in 2004. He then received the M.Sc. (with distinction) in Communication Systems and Signal Processing from the University of Bristol, U.K. and Ph.D. degrees from the University of Edinburgh, U.K. in 2005 and 2008, respectively. From September 2008 through May 2010, he was with the Institute for Circuit Theory and Signal Processing, Munich University of Technology (TUM), Germany working as a Postdoctoral Research Associate. He is currently a Reader (equivalent to Associate Professor) in Multiple-Antenna Systems at Queen's University Belfast, U.K. after holding an Assistant Professor position at Chalmers University of Technology, Sweden. His research interests span signal processing for wireless communications, massive MIMO, hardware-constrained communications, and performance analysis of fading channels.

Jie Yang (yangjie@seu.edu.cn) is currently working towards the Ph.D. degree in communication and information systems with Southeast University, China. Her current research interests include millimeter-wave wireless communication, massive multiple-input multiple-output, and compressed sensing.

Chao-Kai Wen (chaokai.wen@mail.nsysu.edu.tw) received the Ph.D. degree from the Institute of Communications Engineering, National Tsing Hua University, Taiwan, in 2004. He was with Industrial Technology Research Institute, Hsinchu, Taiwan and MediaTek Inc., Hsinchu, Taiwan, from 2004 to 2009. Since 2009, he has been with National Sun Yat-sen University, Taiwan, where he is Professor of the Institute of Communications Engineering. His research interests center around the optimization in wireless multimedia networks.

Feifei Gao [M'09-SM'14] (feifeigao@ieee.org) received the Ph.D. degree from National University of Singapore, Singapore in 2007. He was a Research Fellow with the Institute for Infocomm Research (I2R), A*STAR, Singapore in 2008 and an Assistant Professor with the School of Engineering and Science, Jacobs University, Bremen, Germany from 2009 to 2010. In 2011, he joined the Department of Automation, Tsinghua University, Beijing, China, where he is currently an Associate Professor. Prof. Gao's research areas include communication theory, signal processing for communications, array signal processing, and convex optimizations.

Shi Jin [SM'17] (jinshi@seu.edu.cn) received the Ph.D. degree in communications and information systems from Southeast University, Nanjing, in 2007. From June 2007 to October 2009, he was a Research Fellow with the Adastral Park Research Campus, University College London, London, U.K. He is currently with the faculty of the National Mobile Communications Research Laboratory, Southeast University. His research interests include space-time wireless communications, random matrix theory, and information theory. Dr. Jin and his coauthors received the 2010 Young Author Best Paper Award by the IEEE Signal Processing Society and the 2011 IEEE Communications Society Stephen O. Rice Prize Paper Award in the field of communication theory.

\clearpage
\clearpage
\clearpage
\clearpage

\end{spacing}
\end{document}